\documentclass[twocolumn,showpacs,prl]{revtex4}
\usepackage[latin9]{inputenc}
\usepackage{units}
\usepackage{amsmath}
\usepackage{amssymb}
\usepackage{graphicx}
\usepackage{esint}
\usepackage{braket}
\usepackage{subfigure}

\makeatletter
\@ifundefined{textcolor}{}
{%
 \definecolor{BLACK}{gray}{0}
 \definecolor{WHITE}{gray}{1}
 \definecolor{RED}{rgb}{1,0,0}
 \definecolor{GREEN}{rgb}{0,1,0}
 \definecolor{BLUE}{rgb}{0,0,1}
 \definecolor{CYAN}{cmyk}{1,0,0,0}
 \definecolor{MAGENTA}{cmyk}{0,1,0,0}
 \definecolor{YELLOW}{cmyk}{0,0,1,0}
}

\newcommand{\Yb}{\ensuremath{^{171}\mathrm{Yb}^+~}}
\newcommand{\up}{\ensuremath{\left|\uparrow\right\rangle}}
\newcommand{\down}{\ensuremath{\left|\downarrow\right\rangle}}

\makeatother

\begin{document}

\title{Realization of Geometric Landau-Zener-St\"{u}ckelberg Interferometry}

\author{Junhua Zhang$^{1}$, Jingning Zhang$^{1}$, Xiang Zhang$^1$, and Kihwan Kim$^{1}$}
\email{kimkihwan@gmail.com}

\affiliation{$^{1}$Center for Quantum Information, Institute for Interdisciplinary Information Sciences, Tsinghua University, Beijing 100084, P. R. China}

\date{\today}
\begin{abstract}
We report the first experimental realization of the geometric Landau-Zener-St\"{u}ckelberg (LZS) interferometry proposed by [Phys. Rev. Lett. 107, 207002 (2011)] in a single trapped ion system. Different from a conventional LZS interferometer, the interference fringes of our geometric interferometer originate solely from geometric phase. We also observe the robustness of the interference contrast against noise or fluctuation in the experimental parameters. Our scheme can be applied to other complex systems subject to relatively large errors in system control.
\end{abstract}

\maketitle


Any quantum operation, performed for quantum information processing or the control of quantum dynamics in complex systems, is subject to two primary kinds of errors: (i) random phase errors arising from environmental coupling or (ii) operational errors originating from imperfections or fluctuations in experimental control \cite{nielsen}. Recently there have been extensive theoretical and experimental studies in reducing random phase errors by the use of dynamical decoupling \cite{Uhrig07,Biercuk09,Jiangfeng09} alongside quantum error corrections \cite{ShorA95,Steane96}. In many quantum systems, however, operational errors may play a more significant role in the fidelty of the relevant quantum operations. Composite pulses or adiabatic manipulation with geometric phase have been extensively studied to reach error rates below the fault-tolerant level with reasonable limitations of control in feasible physical systems \cite{Chuang04,Nakahara12}. 

Generally, a quantum system subjected to adiabatic driving acquires a geometric phase (or Berry phase) as well as a dynamic phase. Unlike the dynamic phase, the geometric phase depends solely on the trajectory of the parameters in the Hamiltonian, and thus is stable against certain types of fluctuations, which have been experimentally observed in various systems \cite{Zwanziger03,Wallraff07,Rauch09}. However, the connection of the phase to the Landau-Zener-St\"{u}ckelberg (LZS) interferometry has not yet been experimentally demonstrated \cite{Pekola11}.  

The interferometry is composed of successive Landau-Zener (LZ) transitions, which occur at an avoided crossing of a driven quantum two-level system. The LZS interferometry is similar to the Mach-Zehnder interferometry in optical systems in that the role of the LZ transition corresponds to that of a coherent beam splitter. The interference of the LZS interferometer is governed by the phase difference accumulated by the two instantaneous energy eigenstates between the subsequent transitions. The LZS interference has been observed in diverse physical systems from atomic or optical systems \cite{Woerdman96,Grimm07,Weitz10} to solid state systems \cite{Nori10,Orlando05,Gossard10,Du11}, where all of the observations are dominated by the dynamic phase or a combination of the dynamic and geometric phases \cite{Woerdman96}.

In our paper, we report the first experimental realization of the LZS interferometry controlled exclusively by the geometric phase inspired by the proposal of Ref. \cite{Pekola11} in a single trapped ion system, which is capable of simulating other quantum two-level (qubit) systems. We observe the robustness of the geometric phase against immense operational errors in \emph{all} possible control parameters by artificially introducing noise into the system. Our demonstration of strong immunity sheds light on the possibility of examining the geometric phase in more complex systems which might be subject to large fluctuations in control parameters. Furthermore, our realization contains the basic procedure for the adiabatic quantum simulation and can be extended to investigate and harness the geometric phase in many-body systems for quantum information processing \cite{Duan01,Chen11}. 

The geometric LZS interferometry of a qubit system can be described by the following Hamiltonian,
\begin{eqnarray}
H_{\rm GLZ}\left(t\right)=\frac{\hbar}{2}{\boldsymbol \sigma}\cdot{\mathbf B}_{\rm eff}\left(t\right),
\label{eq:Ham}
\end{eqnarray}
where ${\boldsymbol \sigma}=\left(\sigma_{x},\sigma_{y},\sigma_{z}\right)$ is the vector of Pauli matrices and the effective magnetic field $\mathbf{B}_{\rm eff}\equiv\left(B_{x},B_{y},B_{z}\right)= \left(\Omega\cos\varphi,\Omega\sin\varphi,\Delta\right)$. Here $\Delta$ is the energy splitting of the qubit system at the avoided crossing and $\Omega$ is the driving strength of the system. We denote $\hbar=1$ for convenience.

\begin{figure*}
  \includegraphics[width=2\columnwidth]{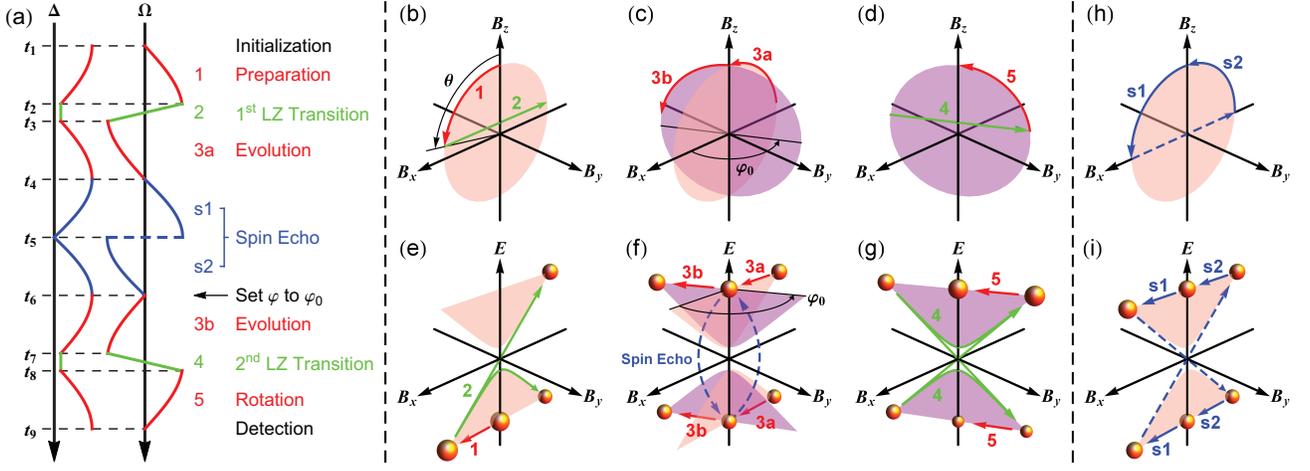}\\
  \caption{(a) The control sequences of $\Delta$ and $\Omega$ in the ${\mathbf B}_{\rm eff} = \left(\Omega \cos \varphi, \Omega \sin \varphi, \Delta \right)$ for the realization of the geometric LZS interferometry. The sequences are composed of: \textbf{1} preparation $[t_1,t_2]$, \textbf{2} the 1$^{\rm st}$ LZ transition $[t_2,t_3]$, \textbf{3a,3b} adiabatic evolutions for aquiring geometric phase $[t_3,t_4]$ $\&$ $[t_6,t_7]$, \textbf{4} the 2$^{\rm nd}$ LZ transition $[t_7,t_8]$, \textbf{5} final rotation $[t_8,t_9]$ and spin echo sequence $[t_4,t_6]$. At $t=t_6$, we change the phase of $\phi$ from 0 to $\phi_{0}$. Adiabatic procedures are noted as red color (1,3a,3b,5). The colors and numbers used in (a) remain consistent in all the other figures. (b-d) The trajectories of the ${\mathbf B}_{\rm eff}$ following the control sequences. (e-g) The description of the geometric LZS interferometry in $E-{\mathbf B}_{\rm eff}^{\bot}$ space, where the hyperbolic curves indicate the adiabatic eigenenergies. In contrast to the standard energy diagram for the LZ transition, the geometric LZS interferometer should be described in 3$D$ space due to the phase information. The volume of the orange spheres corresponds to the population of the adiabatic eigenstates. (h) The trajectories of the ${\mathbf B}_{\rm eff}$ for the adiabatic spin echo sequences. (i) The description of the spin echo sequences $E-{\mathbf B}_{\rm eff}^{\bot}$ space.}\label{fig:ExpSeq}
\end{figure*}

We realize the geometric LZS interferometer in a single \Yb ion as a model qubit. The single \Yb ion is trapped in a four-rods radio frequency trap \cite{Olmschenk07,Zhang12}. We map the two internal levels of the \Yb ion in the $S_{1/2}$ ground state manifold to the qubit states, which is represented by: $\ket{F=1, m_F=0}\equiv\up$ and $ \ket{F=0, m_F=0} \equiv \down$. The energy splitting of the two levels results from the hyperfine interaction and the transition frequency between the $\up$ state and the $\down$ state is $\omega_{\rm HF}= (2 \pi) 12642.821$ MHz. The coherent driving is implemented by the microwave with a frequency detuned by $\Delta(t) = \omega_{\rm HF}-\omega_{\rm M}(t)$ and strength $\Omega$. We control $\Delta$ and $\Omega$ by mixing the microwave signal with the output of an arbitrary wave form generator with 1 GS/s, which is significantly high enough to ignore the sampling effect compared to typical operation time, a few hundred $\mu$s. By going to interaction picture defined by $H_{0}=\frac{\sigma_{z}}{2}\omega_{\rm M}(t)$, we obtain the geometric LZS Hamiltonian, $H_{\rm GLZ}$ (\ref{eq:Ham}), where $\varphi$ is the phase of the microwave source. In experiment, we first apply Doppler cooling and initialize the state to the $\down$ state by the standard optical pumping technique with $99.1\%$ efficiency \cite{Olmschenk07,Zhang12}. At the end of the experimental sequence, we measure the population of the $\up$ state by applying the fluorescent detection scheme \cite{Olmschenk07,Zhang12}.

Fig.~\ref{fig:ExpSeq}(a) shows the sequences of experimental controls in $\Delta$ and $\Omega$ for the geometric LZS interferometry. Fig.~\ref{fig:ExpSeq}(b-e) show the trajectories of the ${\mathbf B}_{\rm eff}$ and Fig.~\ref{fig:ExpSeq}(f-i) show the evolutions of qubit state in $E-{\mathbf B}_{\rm eff}^{\bot}$ space according to changes of parameters in the Hamiltonian, where the hyperbolic curves indicate the adiabatic eigenenergies $E_{\pm}=\pm \sqrt{\Delta_{0}^2+\Omega^2}$. The sequences are composed of the five main procedures and adibatic spin echo: \textbf{1} $[t_1,t_2]$ adiabatic prepartion of the instantaneous ground state of the initial Hamiltonian, \textbf{2} $[t_2,t_3]$ LZ transtion, \textbf{3a,3b} $[t_3,t_4]\&[t_6,t_7]$ adiabatic evolutions to accumulate geometric phase, \textbf{4} $[t_7,t_8]$ LZ transtion, \textbf{5} $[t_8,t_9]$ adiabatic rotation to transfer the final state on the measurement basis $\sigma_z$, and \textbf{S1,S2} $[t_4,t_6]$ spin echo sequence to eliminate the dynamic phase and the Stokes phase. The interference pattern of the geometric phase is observed by measuring the population of the upper eigenstate in the measurement basis.

For the implementation of the geometric LZS interferometer, the adiabatic evolution, where a parameter of a Hamiltonian slowly changes and the system follows the ground state of the instantaneous Hamiltonian, plays crucial role as the main methods for state preparation, state detection, state inversion and geometric phase generation. Therefore, we carefully investigate the validity of adiabaticity in our experimental realization. We also perform the experimental study of the LZ transition, where the separation probability between two instantaneous energy eigenstates is controlled by the sweeping rate in the vicinity of the avoided crossing. Based on the experimental confirmation of the validity of adiabatic evolution and of ability to control LZ transition, we perform the geometric LZS interferometry.

\begin{figure}
  \includegraphics[width=1\columnwidth]{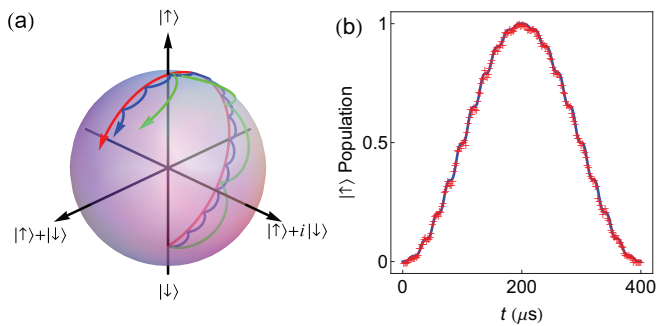}\\
  \caption{(a) The adiabatic evolution of the state. The traces of the system depending on the speeds of changing parameters are shown on the Bloch sphere in case of $\varphi=0$. The red curve is the trace of the perfect adiabatic evolution from the $\down$ state to the ground state of $H_{\rm GLZ}=\frac{\Delta}{2}\sigma_z+\frac{\Omega_i}{2}\sigma_x$. We change $\Delta$ and $\Omega$ as shown in Eq.(\ref{eq:Adiabatic}). The blue and the green curves are the trajectories with speeds $\theta_{f}/T_{a} = \pi /(200\mu$s) and $\pi/(60\mu$s), respectively, with $|\mathbf{B}_{\rm eff}| = (2\pi) 50 $ kHz. (b) The population of the $\up$ state. During the adiabatic rotation, the rotating speed is $\pi/(200\mu{\rm s})$, which is used though out the rest of the paper. The solid line shows numerical estimation of the evolution in $\up$ state population and the $+$ symbols show the experimental data after averaging the measurements of 1500 trials, where the projection error of each point (0.012) is smaller than the size of the symbol. Note that we use the same coventions and perform the same number of trials for one data point throughout this paper.}\label{fig:Adia}
\end{figure}

\textbf{\emph{Adiabatic Evolution}} The adiabatic evolution is used for the sequences of \textbf{1,3a,3b,5} as well as \textbf{S1,S2}. We apply adiabatic evolution to rotate the $\mathbf{B}_{\rm eff}$ field about an axis on the $xy$-plane by changing the amplitude $\Omega$ and the detuning $\Delta$ of the microwave in the following manner,
\begin{eqnarray}
\Omega\left(t\right)=|\mathbf{B}_{\rm eff}| \sin\theta\left(t\right), \Delta\left(t\right)=|\mathbf{B}_{\rm eff}|\cos\theta\left(t\right),
\label{eq:Adiabatic}
\end{eqnarray}
where $|\mathbf{B}_{\rm eff}|$ is the magnitude of the effective magnetic field, $\theta$ is linearly increasing in time $\theta\left(t\right)= \left(\theta_{f}/T_{a}\right)t$. We chose the changing rate $\theta_{f}/T_{a} = \pi/(200\mu$s) for the $|\mathbf{B}_{\rm eff}|= (2 \pi) 50$ kHz, which is small enough to satisfy the adiabaticity. With the rate, the initial ground state $\down$ evolves as a blue curve shown on the Bloch sphere of Fig.~\ref{fig:Adia}(a), where $\varphi=0$. The time-dependency of the populations of the $\up$ state are measured and compared to the numerical calculations [Fig.~\ref{fig:Adia}(b)]. The difference in population between the experimental data and the ideal adiabatic evolution is no more than $5.5\%$. 

\textbf{\emph{LZ Transition}} The LZ transition is used for the sequences of \textbf{2} and \textbf{4}. The LZ transition is investiaged including the time-resolved measurement of the tunneling dynamics similarly to the demonstration with cold atoms \cite{Arimondo09}. LZ tunneling occurs in the vicinity of the avoided crossing and in in the long time limit, the probability transferred to the upper energy eigenstate of the adiabatic basis after the transition is characterized by 
\begin{eqnarray}
P_{\rm LZ}=\exp\left(-\frac{\pi\Delta_{0}^{2}}{2|v|}\right),
\label{eq:LZprob}
\end{eqnarray}
where $v=\left.\frac{\textrm{d}\Omega}{\textrm{d}t}\right|_{\Omega=0}$. In the adiabatic impulse approach, the transition in the adiabatic basis $\left\{ \ket{\psi_{-}}, \ket{\psi_{+}} \right\}$ is described by the evolution matrix $U_{\rm LZ}=U_a N U_b$ with
\begin{eqnarray}
N=\begin{pmatrix}e^{-i\varphi_{S}}\sqrt{1-P_{\rm LZ}} & -\sqrt{P_{\rm LZ}}\\
\sqrt{P_{\rm LZ}} & e^{i\varphi_{S}}\sqrt{1-P_{\rm LZ}}
\end{pmatrix},
\label{eq:ULZ}
\end{eqnarray}
where $\varphi_s$ is the Stokes phase \cite{Kayanuma97} and $U_{a(b)} = \exp\left(i\xi_{a(b)}\sigma_z\right)$, where $\xi_{a(b)}$ is the dynamic phase accumulated after (or before) the LZ transition point.

\begin{figure}
  \includegraphics[width=1\columnwidth]{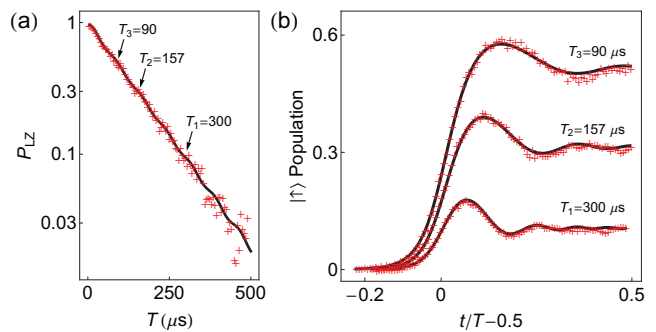}\\
  \caption{(a) LZ tunneling probability $P_{\rm LZ}$ as the function of the total sweeping time $T$ in the linear changes of $\Omega$, where $\Delta_{0} = (2\pi) 8.68$ kHz and $\Omega_{i} = (2\pi) 49.24$ kHz. (b) The LZ transition dynamics for the three exemplary cases: the total durations $T$ are 90 $\mu$s, 157 $\mu$s, and 300 $\mu$s, which provide 0.5, 0.3 and 0.1 tunneling probabilities after the transitions. The oscillatory behaviors near the transition points \cite{Arimondo09,Du11} are clearly observed and precisely agreed with the numerical calculations.}\label{fig:LZprob}
\end{figure}

Experimentally, we use the sequences \textbf{1}, \textbf{2} and \textbf{5} of Fig. \ref{fig:ExpSeq} to study the LZ transtions: \textbf{[Sequence 1]} We prepare the ground state of the Hamiltonian (\ref{eq:Ham}) with $\Omega_{i} = (2\pi) 49.24$ kHz, $\Delta_{0} = (2\pi) 8.68$ kHz and $\varphi=0$ by adiabatically rotating the $\down$ state about $y$-axis; \textbf{[Sequence 2]} We change $\Omega\left(t\right) =\left(1-2\frac{t}{T}\right)\Omega_{i}$ linearly in time; \textbf{[Sequence 5]} At time $t$, we adiabatically bring the state to that in the measurement basis ($\ket{\psi_{-}\left(t\right)} \rightarrow \down, \ket{\psi_{+}\left(t\right)} \rightarrow \up$), which enables us to measure the population of the excited state $\ket{\psi_{+}\left(t\right)}$ by observing the probability of the $\up$ state, which is equivalent to the transition probability. The population of the $\up$ state after the transition versus the duration of the transition is plotted in Fig.~\ref{fig:LZprob}(a), where experimental results and the transition formula of Eq. (\ref{eq:LZprob}) are in precise agreement. We also observe the transient dynamics and oscillatory behavior of LZ tunneling in the vicinity of transition point with various speeds of changing $\Omega$ shown in Fig. ~\ref{fig:LZprob}(b) with improved quantitative agreements compared to the Ref. \cite{Arimondo09,Du11}. Note that there is no fitting parameter at the theoretical expectations in Fig.~\ref{fig:LZprob}, since the parameters are independently measured. 

\begin{figure*}
  \includegraphics[width=2\columnwidth]{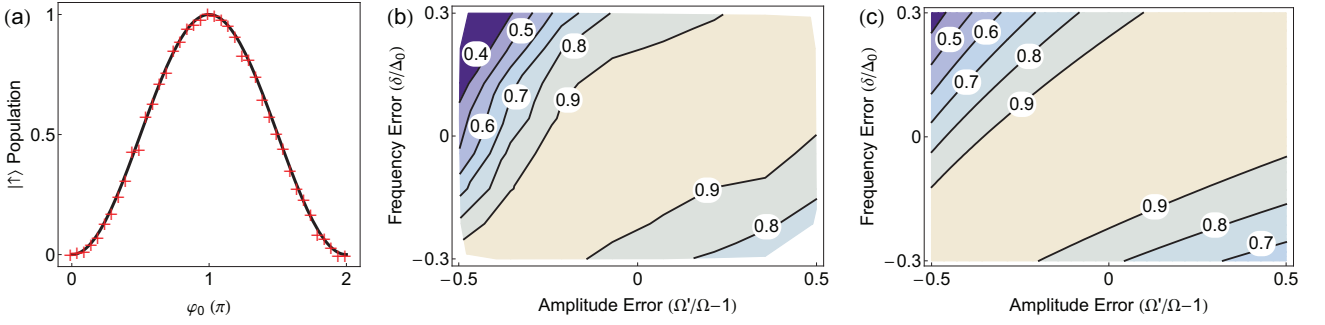}\\
  \caption{(a) The interference pattern of the geometric LZS interferometery. The final $\up$ state populations are only determined by the geometric phase of the rotation angle $\varphi_{0}$. The solid black line comes from the theoretical expectation Eq. (\ref{eq:PGeoLZS}) and $'+'$ symbols are experimental results with the average of 1500 trials. (b) The experimental demonstration and (c) the theoretical estimation of the immunities in the interference contrast against the errors in the amplitude $\Omega$ and the frequency $\Delta$ of $\mathbf{B}_{\rm eff}$. The amplitute and the frequency errors are scaled by the ideal $\Omega$ and the minimum frequency gap $\Delta_{0}$. The experimental figure was constructed by the fringe contrasts of 1139 random choices of the amplitute $\Omega'$ with 1500 repetitions per choice. In our scheme, the constasts of the geometric LZS interferometer are clearly revealed even close to 30$\%$ and 50$\%$ fluctuations of the frequency and amplitude.}\label{fig:FinalResult}
\end{figure*}

\textbf{\emph{Geometric LZS Interferometer}} The sequences for the geometric LZS interferometer is described in Fig. \ref{fig:ExpSeq}(a), where we use the squence \textbf{1},\textbf{2} and \textbf{5} in the same way to the study for the LZ transition as described above. After the squence \textbf{1,2}, the geometric phase is accumulated by the adiabatic evolutions shown in sequence \textbf{3a} and \textbf{3b} combined with the adiabatic spin echo sequence \textbf{S1} and \textbf{S2} in Fig. \ref{fig:ExpSeq}. 

\textbf{[Sequence 3a,3b]} $\&$ \textbf{[Sequence S1,S2]} The acculumated geometric phase of each adiabatic eigenstate $\left\{ \ket{\psi_{-}}, \ket{\psi_{+}} \right\}$ is given by,
\begin{eqnarray}
\gamma_{\pm}=i\int_{t_3}^{t_5}{\bra{\psi_{\pm}\left(t\right)} \frac{d}{dt} \ket{\psi_{\pm}\left(t\right)}} \textrm{d}t
\label{eq:Gphase}
\end{eqnarray}
The difference of geometric phase between upper and lower eigenstates after the evolution is only determined by the rotating angle $\varphi_{0}= \left( \gamma_{+}-\gamma_{-} \right)$ of the ${\mathbf B}_{\rm eff}$ shown in Fig. \ref{fig:ExpSeq}(c), which is independent of the energy difference, the field amplitute change or the duration of the interferometry. The dynamic phase is also acquired in the adiabatic processes. In order to erase the dynamical phase, we adiabatically invert only the state as depicted in Fig. \ref{fig:ExpSeq}(e,i) in the middle of the evolution \cite{Filipp12}. Note that the ${\mathbf B}_{\rm eff}$ is conserved (same before and after the spin echo sequences). The time evolution operation $U_{G}$ for these adiabatic stages including the spin echo sequences is written as,
\begin{eqnarray}
U_{G}=\begin{pmatrix} e^{i\gamma_{-}} & 0\\
0 & e^{i\gamma_{+}}
\end{pmatrix}.
\label{eq:GEvolOperator}
\end{eqnarray}

\textbf{[Sequence 4]} The second LZ transition at $[t_7,t_8]$ in Fig. \ref{fig:ExpSeq}(a) should be described in the basis of the inverted ${\mathbf B}_{\rm eff}$, which results in $\tilde{U}_{\rm LZ}=\sigma_{x} U_{\rm LZ} \sigma_{x}^{-1}$. The final state $\ket{\psi \left(t_{f}\right)}$ after the second LZ transition can be expressed in the following equation,
\begin{eqnarray}
\ket{\psi_{f}}=\tilde{U}_{\rm LZ} U_{G} U_{\rm LZ}\down,
\label{eq:TotalOperator}
\end{eqnarray}
where the initial state is prepared at the instantaneous eigenstate with lower energy of the beginning Hamiltonian. Note that the Stokes phases occurring at the first and second LZ transitions are effectively cancelled out because of the inversion of the state. Therefore, the final result is insensitive to the fluctuation of the Stokes phase as well as the dynamic phase. 

\textbf{[Sequence 5]} At the final rotation $[t_8,t_9]$ in Fig. \ref{fig:ExpSeq}(a), we adiabatically transfer the population of the adiabatic basis to the measurement basis as discussed in the adiabatic process. Finally the population of the $\up$ state after the geometric LZS interferometry can be described by the simple formula, 
\begin{eqnarray}
P_{\up}=2P_{\rm LZ}\left(1-P_{\rm LZ}\right)\left(1-\cos\varphi_0\right).
\label{eq:PGeoLZS}
\end{eqnarray}
In the experiment, we set the transition probability $P_{\rm LZ}=1/2$, which simplifies (\ref{eq:PGeoLZS}) to $P_{\up} =\frac{1}{2} \left(1-\cos\varphi_0\right)$. 

Fig. \ref{fig:FinalResult}(a) clearly shows that the populations of the $\up$ at the end of the interferometry are solely determined by the geometric phase acquired during the adiabatic evolution, which is exactly the rotating angle $\phi_0$. Fig. \ref{fig:FinalResult}(b) and (c) show the immunity of our interferometry to the errors in the amplitude $\Omega$ and the frequency $\Delta$ that are all parameters in the effective field $\mathbf{B}_{\rm eff}$. For the amplitude error scaled by the ideal strength, about $\pm 50\%$ changes reduce the contrast of the interference pattern by 20$\%$, and for the frequency errors relative to the minimum gap $\Delta_{0}$, $\pm 30\%$ offsets decrease the contrast by 20$\%$. Here, we assume $\Omega'$ and $\Delta+\delta$ to be unchanged in a single experiment. In this case, the dynamic phase and Stokes phase are always cancelled out, while the probability of the LZ transition changes between different experiments, which cause the reduction of the interference contrast. We note it has been shown that the geometric phase is robust against the fast fluctuations of the control parameters, where the evolution time is longer than the typical noise correlation time \cite{Rauch09}.

In conclusion, we have realized a clear connection between the geometric phase and the LZS interferometry and have observed the interference of pure geometric phase. We have demonstrated the robustness of the inteference against variations of all the parameters in $\mathbf{B}_{\rm eff}$, which shows the possibility of observing such intereference even in more complex systems, including super-conducting qubit systems \cite{Pekola11,Filipp12}, quantum dot systems \cite{Morton12}, NV center diamond systems \cite{Du11}, etc. Within a trapped ion system, our research makes possible a new direction in performing multi-qubit operations with the geometric phase in internal states \cite{Duan01,Lidar09}, which can provide strong robustness and high fidelity of operation beyond the current methods that use the geometric phase in harmonic oscillators \cite{Wineland03,Roos08}. We also note that the experimental method used in the realization can be directly applied to adiabatic quantum computation. 

This work was supported in part by the National Basic Research Program of China Grant 2011CBA00300, 2011CBA00301, 2011CBA00302, the National Natural Science Foundation of China Grant 61073174, 61033001, 61061130540. KK acknowledges the support from young 1000 plan.

\bibliography{LZSI}
\end{document}